\documentclass{article}
\usepackage{amsfonts}
\usepackage{spconf,amsmath,graphicx}
\input epsf
\title{\vspace{-1em} Adaptive Reduced-Rank MBER Linear Receive Processing for Large Multiuser MIMO Systems \vspace{-0.5em}}
\name{Yunlong Cai{\small $~^{\#1}$} and Rodrigo C. de Lamare{\small
$~^{*2}$} \vspace{-1em}}
\address{$^{\#}$\,Department of ISEE, Zhejiang University,  Hangzhou 310027, China \\
 $^{*}$\,Department of Electronics, University of York, York, UK, YO10 5DD\\   $^{1}$\,ylcai@zju.edu.cn, $^{2}$\,rcdl500@ohm.york.ac.uk \vspace{-1em}}
\begin{document}
\ninept

\maketitle
\begin{abstract}
In this work, we propose a novel adaptive reduced-rank strategy
based on joint interpolation, decimation and filtering  (JIDF)  for
large multiuser multiple-input multiple-output (MIMO) systems. In
this scheme, a reduced-rank framework is proposed for linear receive
processing and multiuser interference suppression according to the
minimization of the bit error rate (BER) cost function. We present a
structure with multiple processing branches that performs
dimensionality reduction, where each branch contains a group of
jointly optimized interpolation and decimation units, followed by a
linear receive filter. We then develop stochastic gradient (SG)
algorithms to compute the parameters of the interpolation and
receive filters along with a low-complexity decimation technique.
Simulation results are presented for time-varying environments and
show that the proposed MBER-JIDF receive processing strategy and
algorithms achieve a superior performance to existing methods at a
reduced complexity.
\end{abstract}

\section{Introduction}
\label{sec:intro}

Large  MIMO systems have received significant attention in the
recent years since they can substantially increase the system
capacity and improve the quality and reliability of wireless links
\cite{largemimo}. Different configurations have been investigated
for large MIMO systems, such as distributed and centralized MIMO
schemes. Key applications of these systems include wireless
cellular, local area \cite{marzetta,li,mohammed} and multi-beam
satellite networks \cite{arnau}. The problem of detecting a desired
user in a large multiuser MIMO system presents many signal
processing challenges including the need for algorithms with the
ability to process large-dimensional received data, fast and
accurate adjustment of parameters, scalable computational complexity
and the development of cost-effective interference mitigation
schemes.

In this context, reduced-rank signal processing is a key tool for
large systems which can provide faster training, a better tracking
performance and an increased robustness against interference as
compared to standard methods. A number of reduced-rank techniques
have been developed to design the dimensionality reduction matrix
and the reduced-rank receive filter
\cite{eign}-\cite{delamaretvt11}. Among the first schemes are the
eigendecomposition-based (EIG) algorithms \cite{eign}, \cite{eign2}
and the multistage Wiener filter (MWF) investigated in
\cite{MWF2}-\cite{ccmmwf}.
EIG and MWF  have faster convergence speed  compared to the full
rank adaptive algorithms  with a much smaller filter size, but their
computational complexity is high. A strategy based on the joint and
iterative optimization (JIO) of a subspace projection matrix and a
reduced-rank filter has been reported in
\cite{delamarespl07,jidf,delamaretvt10,delamaretvt11}.

However, most of the contributions to date are either based on the
minimization of the mean square error (MSE) and/or the minimum
variance criteria \cite{eign}-\cite{delamaretvt11}, which are not
the most appropriate  metric from a performance viewpoint in digital
communications. Design approaches that can minimize the bit error
rate (BER) have been reported in \cite{mber2,delamare_mber,mber3,
euroship,mberjio} and are termed adaptive MBER techniques. The work
in \cite{mber3} appears to be the first approach to combine a
reduced-rank algorithm with the BER criterion. However, the scheme
is a hybrid  between an EIG or an MWF approach, and a BER scheme in
which only the reduced-rank filter is adjusted in an MBER fashion.
Moreover, the existing works on MBER techniques have not addressed
the key problem of performance degradation experienced when the
filters become larger and their performance converges gradually to
MSE-based techniques.


In this work, we propose an adaptive reduced-rank linear receive
processing strategy based on joint interpolation, decimation and
filtering (JIDF) for large multiuser MIMO systems. The proposed
scheme employs a multiple-branch framework which adaptively performs
dimensionality reduction using a set of jointly optimized
interpolation and decimation units, followed by receive filtering
according to the BER cost function. The dimensionality reduction is
optimized at each time instant by selection of the interpolation
filter and the decimation pattern with the best performance. After
dimensionality reduction, a linear receive filter with reduced
dimension designed using the BER criterion is applied to suppress
the multiuser interference and estimate the data symbols. We devise
stochastic gradient (SG) algorithms to compute the parameters of the
interpolation and receive filters along with a low-complexity
decimation technique. A unique feature of our scheme is that all
component filters have a small number of parameters and can take
full advantage of the MBER adaptation. Simulation results show that
the proposed MBER-JIDF receive processing strategy and algorithms
have a superior performance to existing techniques at a reduced
complexity.


\section{System Model and Problem Statement}
\label{Section2:system}

Let us consider the uplink of an uncoded synchronous multiuser MIMO
system with $K$ users and one base station (BS)
\cite{delamaretc}-\cite{peng_twc}, where each user is equipped with
$N_U$ antennas and the BS is equipped with $M$ uncorrelated receive
antennas and $KN_{U}< M$. We assume that the channel is a MIMO
time-varying flat fading channel.
%
The
$M$-dimensional received vector is given by
\begin{equation}
\begin{split}
\mathbf {r}(i) & = \sum_{k=1}^{K}A_{k}\mathbf{H}_{k}(i) \mathbf
{b}_{k}(i) + \mathbf {n}(i),
\end{split}
\end{equation}
where  $\mathbf{b}_{k}(i)=[b_{k,1}(i) \ldots b_{k,n}(i) \ldots b_{k,N_U}(i)]^T$ is a
$N_U \times 1$ symbol vector of user $k$ corresponding to the $i$-th
time instant, $n=1,\ldots, N_{U}$, and the amplitude of user $k$ is $A_{k}$,
$k=1,\ldots,K$. The $M \times N_U$ matrix $\mathbf{H}_{k}(i)$ is the
channel matrix of user $k$, which is given by
 \begin{equation}
 \mathbf {H}_{k}(i)=[{\mathbf h}_{k,1}(i) \ldots {\mathbf
 h}_{k,f}(i) \ldots
{\mathbf h}_{k,M}(i)]^{T},
 \end{equation}
 %
where the $N_U \times 1$ channel vectors ${\bf h}_{k,f}(i)$, for
$f=1,\ldots, M$, consist of independent and identically distributed
complex Gaussian variables with zero mean and unit variance,
$\mathbf {n}(i) = [n_{1}(i) \ldots n_{M}(i)]^T$ is the complex
Gaussian noise vector with zero mean and $E[\mathbf {n}(i)\mathbf
{n}^{H}(i)] = \sigma^2 \mathbf {I}$, where $\sigma^2$ is the noise
variance,  $(.)^T$ and $(.)^H$ denote transpose and Hermitian
transpose, respectively.


In the following, we explain the design of reduced-rank receive
processing schemes which minimize the BER. In a reduced-rank
algorithm, an $M \times D$ subspace projection matrix
$\mathbf{S}_{D}$ is applied to the received data to extract the most
important information of the data by performing dimensionality
reduction, where $1 \leq D \leq M$. A $D \times 1$ projected
received vector is obtained as $\mathbf {\bar{r}}(i)=\mathbf
{S}^{H}_{D}\mathbf {r}(i)$,
where it is the input to a $D\times 1$ filter $\mathbf
{\bar{w}}$.
The
filter output is given by $\bar{x}_{k,n}(i)=\mathbf {\bar{w}}^{H}\mathbf
{\bar{r}}(i)=\mathbf {\bar{w}}^{H} \mathbf {S}^{H}_{D}\mathbf
{r}(i)$. Assuming that we use binary signalling,
%
the estimated symbol of user $k$ is given by $\hat{b}_{k,n}(i)=\textrm {sign}\{\Re[\mathbf{\bar{w}}^{H}\mathbf{\bar{r}}(i)]\}$,
where the operator $\Re[.]$ retains the real part of the
argument and $\textrm {sign} \{.\}$ is the signum function.
The probability of error for user $k$ is given by
\begin{equation}
\begin{split}
P_{e} &=\int^{0}_{-\infty}f(\tilde{x}_{k,n}) d\tilde{x}_{k,n}= Q \bigg( \frac{\textrm{sign} \{ b_{k,n}(i)\}\Re[\bar{x}_{k,n}(i)]}{\rho (\mathbf{\bar{w}}^{H}\mathbf{S}_{D}^{H}\mathbf{S}_{D}\mathbf{\bar{w}})^{\frac{1}{2}}} \bigg),\label{eq:proberror}
\end{split}
\end{equation}
where $\tilde{x}_{k,n}=\textrm{sign} \{
b_{k,n}(i)\}\Re[\bar{x}_{k,n}(i)]$ denotes a random variable, $f(\tilde{x}_{k,n})$ is the single
point kernel density estimate \cite{mber2} which is given by
\begin{equation}
\begin{split}
f(\tilde{x}_{k,n})=&\frac{1}{\rho \sqrt{2\pi \mathbf{\bar{w}}^{H}\mathbf{S}_{D}^{H}\mathbf{S}_{D}\mathbf{\bar{w}} }}
\exp \bigg(
\frac{-(\tilde{x}_{k,n}-\textrm{sign}  \{ b_{k,n}\}\Re[\bar{x}_{k,n}])^2}{2\mathbf{\bar{w}}^{H}\mathbf{S}_{D}^{H}\mathbf{S}_{D}\mathbf{\bar{w}}\rho^2}\bigg),
\end{split}
\end{equation}
where $\rho$ is the radius parameter of the kernel density estimate,
$Q(.)$ is the Gaussian error function.
The problem
we are interested in solving is how to devise a cost-effective
algorithm to adjust the parameters of $\mathbf
{S}_D$ and $\mathbf{ \bar{w}}$ based on minimizing the probability of error with reduced length component filters.

\section{Proposed MBER-JIDF Reduced-Rank Linear Receive Processing Scheme}
\label{Section4:proposedVFF}

In this section, we detail the proposed MBER reduced-rank linear
receive processing scheme based on joint interpolation, decimation
and filtering, which comes from two observations. The first is that
rank reduction can be performed by reconstructing new samples with
interpolators and eliminating (decimating) samples that are not
useful in the filtering process \cite{delamaretvt10}. The second
comes from the structure of the dimensionality reduction matrix,
whose columns are a set of vectors formed by the interpolators and
decimators.

%
%

\subsection{Overview of the MBER-JIDF Scheme}

\begin{figure}[!hhh]
\centering \scalebox{0.45}{\includegraphics{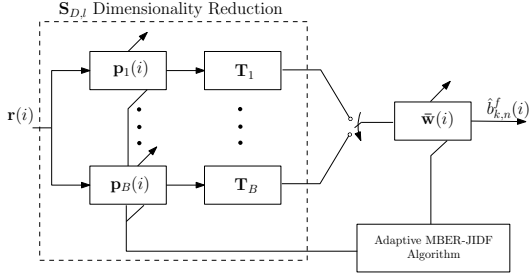}}
\vspace{-1em}\caption{Structure of the proposed MBER-JIDF scheme}
\label{fig:jidfmber}
\end{figure}

We design the subspace projection matrix $\mathbf{S}_{D}$ by
considering interpolation and decimation. In this case, the receive
filter length is substantially reduced, which results in
significantly reduced computational complexity and very fast
training for large MIMO systems. The proposed
MBER-JIDF scheme for the $n$-th symbol of  the $k$-th user is depicted in Fig.
\ref{fig:jidfmber}. The $M \times1$ received vector $\mathbf{r}(i)$
is processed by a framework with $B$ branches, where each branch
contains an interpolator and a decimation unit, followed by a reduced-rank receive filter. In the $l$-th branch, the received vector is
operated by the interpolator
$\mathbf{p}_{l}(i)=[p_{1,l}(i),\ldots, p_{I,l}(i)]^{T}$ with
filter length $I$, $I < M$, the output of the interpolator of the
$l$-th branch is expressed by
\begin{equation}
\mathbf{\tilde{r}}_{l}(i)=\mathbf{P}^{H}_{l}(i)\mathbf{r}(i)
\end{equation}
where the $M\times M$ Toeplitz convolution matrix $\mathbf{P}_{l}(i)$ is given by
\begin {displaymath}
\mathbf {P}_{l}(i)= \left( \begin{array}{cccc} p_{1,l}(i) & 0 & \ldots & 0 \\
\vdots & p_{1,l}(i)  & \ldots & 0\\
p_{I,l}(i) & \vdots & \ldots &0 \\
0 & p_{I,l}(i) & \ldots & 0 \\
0 & 0 & \ddots & 0\\
\vdots & \vdots & \ddots & \vdots\\
0 & 0 & \ldots &  p_{1,l}(i)
\end{array} \right).
\end{displaymath}
In order to facilitate the description of the scheme, we introduce
an alternative way to represent the vector
$\mathbf{\tilde{r}}_{l}(i)$,
\begin{equation}
\mathbf{\tilde{r}}_{l}(i)=\mathbf{P}^{H}_{l}(i)\mathbf{r}(i)=\mathbf{R}^{'}(i)\mathbf{p}^{*}_{l}(i)
\end{equation}
where the $M\times I$ matrix $\mathbf{R}^{'}(i)$ with the samples of
$\mathbf{r}(i)=[r_{0}(i),\ldots, r_{M-1}(i)]^{T}$ has a Hankel
structure \cite{hankel} given by
\begin {displaymath}
\mathbf{R}^{'}(i)= \left( \begin{array}{cccc} r_{0}(i) & r_{1}(i) & \ldots & r_{I-1}(i) \\
\vdots & \vdots & \ldots & \vdots \\
r_{M-1}(i) & r_{M-I+1}(i) & \ldots & r_{M-1}(i)\\
r_{M-I+1}(i) & r_{M-I+2}(i) & \ddots & 0\\
\vdots & \vdots & \ddots & \vdots\\
r_{M-2}(i) & r_{M-1}(i) &0 & 0\\
r_{M-1}(i) & 0 & 0 & 0
\end{array} \right).
\end{displaymath}
The dimensionality reduction is performed by a decimation unit with
$D\times M$ decimation matrices $\mathbf{T}_{l}$ that projects
$\mathbf{\tilde{r}}_{l}(i)$ onto $D\times 1$ vectors
$\mathbf{\bar{r}}_{l}(i)$ with $l=1,\ldots, B$, where $D$ is the
rank. The $D \times 1$ vector $\mathbf{\bar{r}}_{l}(i)$ for the
$l$-th branch is given by
\begin{equation}
\mathbf{\bar{r}}_{l}(i)=\underbrace{\mathbf{T}_{l}\mathbf{P}^{H}_{l}(i)}_{\mathbf{S}_{D,l}(i)}\mathbf{r}(i)=\mathbf{T}_{l}\mathbf{\tilde{r}}_{l}(i)=\mathbf{T}_{l}\mathbf{R}^{'}(i)\mathbf{p}^{*}_{l}(i)
\end{equation}
where $\mathbf{S}_{D,l}(i)$ denotes the equivalent subspace
projection matrix corresponding to the $l$-th branch.  The output of
the reduced-rank receive filter $\mathbf{\bar{w}}(i)$ corresponding to the $l$-th branch
is given by
$\bar{x}^{l}_{k,n}(i)=\mathbf{\bar{w}}^{H}(i)\mathbf{\bar{r}}_{l}(i)$,
which is used in the minimization of the error probability for
branch $l$. The hard decision for the $l$-th branch is given by
$\hat{b}^{l}_{k,n}(i)=\textrm
{sign}\{\Re[\mathbf{\bar{w}}^{H}\mathbf{\bar{r}}_{l}(i)]\}$.
The proposed scheme employs $B$ parallel branches of interpolators and
decimators.
The optimum branch  is selected according to
\begin{equation}
l_{opt}=\arg \min_{1\leq l \leq B} P^{(l)}_{e}, ~{\rm where}~
P^{(l)}_{e}=Q \bigg( \frac{\textrm{sign} \{
b_{k,n}(i)\}\Re[\bar{x}^{l}_{k,n}(i)]}{\rho
(\mathbf{\bar{w}}^{H}\mathbf{S}^{H}_{D,l}\mathbf{S}_{D,l}\mathbf{\bar{w}})^{\frac{1}{2}}}
\bigg).\label{eq:errorpl}
\end{equation}
The output of the scheme is given by $\hat{b}^{f}_{k,n}(i)=\textrm
{sign}\{\Re[\mathbf{\bar{w}}^{H}\mathbf{\bar{r}}_{l_{opt}}(i)]\}$.

\subsection{Design of the Decimation Unit}

In this work, the elements of the decimation matrix only take the value $0$ or $1$. This corresponds to the decimation unit simply keeping or discarding the samples.
The optimal decimation scheme exhaustively explores all possible patterns which select $D$ samples out of $M$ samples.
In this case, the scheme can be viewed as a combinatorial problem and the total number of patterns is $B=M (M-1)\ldots(M-D+1)$.

However, the optimal decimation scheme is too complex for practical use. We introduce a low-complexity suboptimal method to generate the decimation matrix.
It employs a structure formed in the following way
\begin {displaymath}
 \mathbf{T}_{l}=\left[ \begin{array}{cccc} \mathbf{t}_{l,1} &  \mathbf{t}_{l,2}  & \ldots  & \mathbf{t}_{l,D}
 \end{array} \right]^{T}
\end{displaymath}
where the $M\times 1$ vector $\mathbf{t}_{l,d}$ denotes the $d$-th basis vector of the $l$-th decimation unit, $d=1,\ldots, D$, $l=1,\ldots, B$, and
its structure is given by
\begin{equation}
\mathbf{t}_{l,d}= [\underbrace{0, \ldots, 0}_{q_{l,d}}, 1, \underbrace{0, \ldots, 0}_{M-q_{l,d}-1}]^{T}
\end{equation}
where $q_{l,d}$ is the number of zeros before the nonzero element.
Note that it is composed of a single $1$ and $M-1$ $0$s. We set the
value of $q_{l,d}$ in a deterministic way which can be expressed as
$q_{l,d}=\lfloor{\frac{M}{D}\rfloor}\times(d-1)+(l-1)$.
The simulation results will show that the proposed reduced-rank
scheme with the suboptimal decimation unit design method works very well.
In the following section, we will introduce the proposed adaptive
algorithms for the interpolator filter $\mathbf{p}_{l}(i)$ and the
reduced-rank receive filter $\mathbf{\bar{w}}(i)$.

\section{Proposed Adaptive  Algorithms}

In this section, we develop the MBER based adaptive SG algorithms to
update the interpolator and the reduced-rank filters for each
branch. We then provide a computational complexity analysis of the
proposed and conventional adaptive reduced-rank algorithms.


\subsection{Adaptive MBER-JIDF Algorithms}

Firstly, we derive the gradient terms for the reduced-rank filter and the interpolation vector.
By taking the gradient of (\ref{eq:errorpl})
with respect to $\mathbf {\bar{w}}^{*}$ and after further
mathematical manipulations we obtain
\begin{equation}
\begin{split}
\frac{\partial P^{(l)}_{e}}{\partial \mathbf{\bar{w}}^{*}} &
=\frac{-\exp \bigg(\frac{-|\Re[\bar{x}^{l}_{k,n}(i)]|^2}{2\rho^2 \mathbf {\bar{w}}^{H}\mathbf {S}_{D,l}^{H}\mathbf {S}_{D,l}
\mathbf {\bar{w}}} \bigg)\textrm{sign} \{ b_{k,n}(i)\}}{2\sqrt{2\pi}\rho}
\\ &\quad\times \bigg( \frac{\mathbf {S}^{H}_{D,l}\mathbf {r}}{(\mathbf {\bar{w}}^{H}
\mathbf {S}_{D,l}^{H} \mathbf {S}_{D,l} \mathbf
{\bar{w}})^{\frac{1}{2}}}-\frac{\Re[\bar{x}^{l}_{k,n}(i)] \mathbf
{S}^{H}_{D,l} \mathbf {S}_{D,l} \mathbf {\bar{w}}}{( \mathbf
{\bar{w}}^{H} \mathbf {S}_{D,l}^{H}\mathbf {S}_{D,l} \mathbf
{\bar{w}})^{\frac{3}{2}}}\bigg)\label{eq:proberror2}.
\end{split}
\end{equation}
To derive the gradient terms for the interpolator
$\mathbf{p}_{l}(i)$, we need to express the output of the $l$-th
branch $\bar{x}^{l}_{k,n}(i)$ as a function of $\mathbf{p}_{l}(i)$,
which is given by
\begin{equation}
\bar{x}^{l}_{k,n}(i)=\mathbf{\bar{w}}^{H}(i)\mathbf{T}_{l}(i)\mathbf{R}^{'}(i)\mathbf{p}^{*}_{l}(i)=\mathbf{p}^{H}_{l}(i)\mathbf{u}(i)
\end{equation}
 where $\mathbf{u}(i)=\mathbf{R}^{'T}(i)\mathbf{T}^{T}_{l}(i)\mathbf{\bar{w}}^{*}(i)$ is an $I\times 1$ vector.
 We let $\mathbf{u}(i)=[u_{1}(i),\ldots, u_{I}(i)]^{T}$ and rewrite the error probability cost function $P^{(l)}_{e}$ as follows
 \begin{equation}
 P^{(l)}_{e}=Q \bigg( \frac{\textrm {sign}\{b_{k,n}(i)\}\Re[p_{1,l}u_{1}+p_{2,l}u_{2}+\ldots+p_{I,l}u_{I}]}{\rho \sqrt{g(p_{1,l}, p_{2,l}, \ldots, p_{I,l})}} \bigg)
 \end{equation}
 where the function $g(p_{1,l}, p_{2,l}, \ldots, p_{I,l})$ is given by
 \begin{equation}
 \begin{split}
 g(p_{1,l}, p_{2,l}, \ldots, p_{I,l})&=\bar{w}_{1}(p_{1,l}p^{*}_{1,l}+\ldots+p_{\phi_{1},l}p^{*}_{\phi_{1},l})\bar{w}^{*}_{1}\\&
 \quad+\bar{w}_{2}(p_{1,l}p^{*}_{1,l}+\ldots+p_{\phi_{2},l}p^{*}_{\phi_{2},l})\bar{w}^{*}_{2}+\ldots\\&
 \quad+\bar{w}_{D}(p_{1,l}p^{*}_{1,l}+\ldots+p_{\phi_{D},l}p^{*}_{\phi_{D},l})\bar{w}^{*}_{D}
\end{split}
 \end{equation}
 where $\phi_{d}$ denotes the number of nonzero elements for row $d$ in the $D\times I$ matrix $\mathbf{T}_{l}(i)\mathbf{R}^{'}(i)$, $1\leq d\leq D$,   $I= \phi_{1} \geq \phi_{2} \geq \ldots \geq \phi_{D} \geq 1$.
 Note that $g(p_{1,l}, p_{2,l}, \ldots, p_{I,l})=\mathbf {\bar{w}}^{H}
\mathbf {S}_{D,l}^{H} \mathbf {S}_{D,l} \mathbf
{\bar{w}}$, and we define $\mathbf{\bar{w}}=[\bar{w}_{1},\ldots, \bar{w}_{D}]^{T}$.
 By taking the gradient with respect to each element $p^{*}_{j,l}$ in vector $\mathbf{p}_{l}(i)$, $j=1,\ldots, I$, we obtain
 \begin{equation}
 \begin{split}
 \frac{\partial P^{(l)}_{e}}{\partial p^{*}_{j,l}} &
 =\frac{-\exp \bigg(\frac{-|\Re[\bar{x}^{l}_{k,n}(i)]|^2}{2\rho^2 g(p_{1,l}, \ldots, p_{I,l})} \bigg)\textrm{sign} \{ b_{k,n}(i)\}}{2\sqrt{2\pi}\rho}
\\ &\times \bigg( \frac{u_{j}(i)}{g^{\frac{1}{2}}(p_{1,l},  \ldots, p_{I,l})}-\frac{\Re[\bar{x}^{l}_{k,n}] (|\bar{w}_{1}|^2+\ldots+|\bar{w}_{\psi_{j}}|^2)p_{j,l} }{ g^{\frac{3}{2}}(p_{1,l},  \ldots, p_{I,l})}\bigg)\label{eq:proberror22},
\end{split}
 \end{equation}
 where $\psi_{j}$ denotes the number of nonzero elements for column $j$ in the $D\times I$ matrix $\mathbf{T}_{l}(i)\mathbf{R}^{'}(i)$, $1\leq j\leq I$, $D = \psi_{1} \geq \psi_{2} \geq \ldots \geq \psi_{I} \geq 1$. We stack the $I$ elements $\frac{\partial P^{(l)}_{e}}{\partial p^{*}_{j,l}}$ and obtain an $I\times 1$ gradient vector as $ \mathbf{v}_{l}=\big[\frac{\partial P^{(l)}_{e}}{\partial p^{*}_{1,l}}, \frac{\partial P^{(l)}_{e}}{\partial p^{*}_{2,l}}, \ldots, \frac{\partial P^{(l)}_{e}}{\partial p^{*}_{I,l}}\big]^{T}$.

The interpolator and the reduced-rank
receive filters  are jointly optimized
according to the BER criterion. The algorithm has been devised to
start its operation in the training (TR) mode, and then to switch to
the decision-directed (DD) mode. The proposed SG algorithms are
obtained by substituting the gradient terms (\ref{eq:proberror2})
and (\ref{eq:proberror22}) in the expressions $\mathbf{
\bar{w}}(i+1)=\mathbf{ \bar{w}}(i)-\mu_{w}\frac{\partial
P_{e}}{\partial \mathbf{ \bar{w}}^{*}}$ and $\mathbf{
p}_{l}(i+1)=\mathbf{ p}_{l}(i)-\mu_{p}\mathbf{v}_{l}$
\cite{haykin} subject to the constraint of $\mathbf{
\bar{w}}^{H}(i)\mathbf{ S}_{D,l}^{H}(i)
\mathbf{ S}_{D,l}(i)\mathbf{ \bar{w}}(i)=g(p_{1,l}, p_{2,l}, \ldots, p_{I,l})=1$. 
%
%
%
At each time instant, the weights of the two
quantities of branch $l$ are updated in an alternating way by using the following
equations
\begin{equation}
\begin{split}
\mathbf {\bar{w}}(i+1)&=\mathbf{
\bar{w}}(i)+\mu_{w}\frac{\exp
\bigg(\frac{-|\Re[\bar{x}^{l}_{k,n}(i)]|^2}{2\rho^2} \bigg)\textrm{sign}
\{ b_{k,n}(i)\}}{2\sqrt{2\pi}\rho}
\\&\quad\times \big( \mathbf{ S}^{ H}_{D,l}(i)\mathbf{ r}(i)-\Re[\bar{x}^{l}_{k,n}(i)]
\mathbf{ S}^{ H}_{D,l}(i)\mathbf{ S}_{ D,l}(i)\mathbf{
\bar{w}}(i)\big)\label{eq:proberror6}
\end{split}
\end{equation}
\begin{equation}
\begin{split}
\mathbf{p}_{l}(i+1)=\mathbf{ p}_{l}(i)-\mu_{p}\times\bigg[\frac{\partial P^{(l)}_{e}}{\partial p^{*}_{1,l}}, \frac{\partial P^{(l)}_{e}}{\partial p^{*}_{2,l}}, \ldots, \frac{\partial P^{(l)}_{e}}{\partial p^{*}_{I,l}}\bigg]^{T}
\label{eq:proberror7}
\end{split}
\end{equation}
where each element in the gradient vector is given by
\begin{equation}
\begin{split}
 \frac{\partial P^{(l)}_{e}}{\partial p^{*}_{j,l}} &
 =\frac{-\exp \bigg(\frac{-|\Re[\bar{x}^{l}_{k,n}(i)]|^2}{2\rho^2 } \bigg)\textrm{sign} \{ b_{k,n}(i)\}}{2\sqrt{2\pi}\rho}
\\ &\quad\times \big( u_{j}(i)-\Re[\bar{x}^{l}_{k,n}(i)] (|\bar{w}_{1}|^2+\ldots+|\bar{w}_{\psi_{j}}|^2)p_{j,l} \big),
\end{split}
\end{equation}
where $\mu_{w}$ and $\mu_{p}$ are the step-size values. Expressions
(\ref{eq:proberror6}) and (\ref{eq:proberror7}) need initial values,
$\mathbf {\bar{w}}(0)$ and $\mathbf{p}_{l}(0)$, and we scale the
interpolation vector by  $\mathbf{ p}_{l}\leftarrow \frac{\mathbf{
p}_{l}}{\sqrt{\mathbf{ \bar{w}}^{H}\mathbf{ S}_{D,l}^{H}\mathbf{
S}_{D,l}\mathbf{ \bar{w}}}}$ at each iteration.  The scaling has an
equivalent performance to using a constrained optimization with
Lagrange multipliers although it is computationally simpler.  The
proposed MBER-JIDF algorithm are summarized in Table
\ref{tab:table1}.

 \begin{table}[h]
\centering
 \caption{\normalsize Proposed adaptive MBER-JIDF algorithm.} {
\begin{tabular}{ll}
\hline
 $1$ & Set step-size values $\mu_{w}$ and $\mu_{S_{D}}$ and the no. of branches $B$.\\
 $2$ & Initialize $\mathbf {\bar{w}}(0)$ and $\mathbf {p}_{l}(0)$. Set $\mathbf{T}_{1},\ldots, \mathbf{T}_{B}$.\\
 $3$ &   for each time instant $i$ do \\
 $4$ & ~~~ for $l$ from $1$ to $B$ do\\
 $5$ & ~~~~~  Update  $\mathbf {p}_{l}(i+1)$ based on $\mathbf{T}_{l}$ and    (\ref{eq:proberror7}). \\
 $6$ & ~~~~~  Scale the vector $\mathbf {p}_{l}$ using $\mathbf {p}_{l} \leftarrow \frac{\mathbf {p}_{l}}{\sqrt{\mathbf {\bar{w}}^{H}\mathbf {S}^{H}_{D,l}\mathbf {S}_{D,l}\mathbf {\bar{w}}}}$.\\
 $6$ & ~~~ end  \\
 $7$ & Select the optimal branch
     based on (\ref{eq:errorpl}). \\
     &  Generate the estimated symbol.\\
 $8$ &  Update $\mathbf{\bar{w}}(i+1)$ based on the selected branch and (\ref{eq:proberror6}) \\
\hline
\label{tab:table1}
\end{tabular}
}
\end{table}
\vspace{-1em}

\subsection{Computational Complexity}
In Table \ref{tab:table2}, we show the number of additions and
multiplications of the proposed MBER-JIDF algorithm, the existing
adaptive reduced-rank algorithms, the adaptive least-mean square
(LMS) \cite{haykin} and the SG full-rank algorithm based on the BER
criterion \cite{mber2}. In the case of large MIMO systems,  the
parameters $D$, $I$ and $B$ are chosen much smaller than $M$, which
results in a substantial complexity saving.
In particular, for a configuration with $M=40$, $I=D=8$ and $B=4$,
the numbers of multiplications and additions for the proposed
algorithm are upper bounded by $1825$ and $1595$, respectively. For
the MWF-MBER algorithm they are $15594$ and $11857$, respectively.
Compared to the existing reduced-rank algorithms, the MBER-JIDF
algorithm reduces the computational complexity significantly.

\vspace{-1em}
\begin{table}[h]
\centering%
\caption{\normalsize Computational complexity of Algorithms.} {
\begin{tabular}{ccc}
\hline \rule{0cm}{2.5ex}&  \multicolumn{2}{c}{Number of operations
per  symbol} \\ \cline{2-3}
Algorithm & {Multiplications} & {Additions} \\
\hline
\emph{\small  Full-Rank-LMS}  & {\small $2M+1$} & {\small $2M$}  \\
\emph{\small  Full-Rank-MBER}  &  {\small $4M+1$} & {\small $4M-1$}  \\
\emph{\small   EIG} \cite{eign2} &   {\small $O(M^3)$} &  {\small $O(M^3)$}  \\
\emph{\small  MBER-MWF} \cite{mber3} &  {\small $(D+1)M^2$} &  {\small $(D-1)M^2$} \\
                       &   {\small $+(3D+1)M$}  &  {\small $+(2D-1)M$}          \\
                       &      {\small $+M+3D+10$}           &      {\small $+M+2D+1$}              \\
\emph{\small  MBER-JIDF} & {\small $MDB+DB$}  & {\small $MDB+IB$} \\
                       &        {\small $+7IB+4D+1$}             &     {\small $-B+4D-1$}            \\
                       &        {\small $+\sum_{l}\sum_{j}\psi_{j}$}                                &        {\small $+\sum_{l}\sum_{j}\psi_{j}$}               \\
\hline
\label{tab:table2}
\end{tabular}
}
\end{table}
\vspace{-2.5em}

\section{Simulations}
\label{Section6:simulations}

In this section, we evaluate the  performance of the proposed
MBER-JIDF reduced-rank algorithm and compare it with existing
full-rank and reduced-rank algorithms. Monte-carlo simulations are
conducted to verify the effectiveness of the MBER-JIDF adaptive
reduced-rank SG algorithms.
The number of receive antennas at the BS is $M=40$. The number of
antennas per user is $N_{U}=2$. The coefficients of the channel
matrix ${\bf H}_{k}(i)$ are computed according to Clarke's model
\cite{rappaport}. We have optimized the step sizes of each branch of
the MBER-JIDF adaptive reduced-rank SG algorithms with the following
rules, $\mu_{w/p}(i+1)=\bigg[ \delta_{1} \mu_{w/p}(i)+
\delta_{2}\times Q\bigg(\frac{\textrm
{sign}\{\hat{b}^{l}_{k,n}(i)\}\Re[\bar{x}^{l}_{k,n}(i)]}{\rho}\bigg)
\bigg]^{\mu^{+}}_{\mu^{-}}$,
where $[.]^{\mu^{+}}_{\mu^{-}}$ denotes the truncation to the limits
of a range. We tuned $\delta_{1}=0.99$, $\delta_{2}=1\times
10^{-4}$, $\mu^{+}=1\times10^{-2}$ and $\mu^{-}=1\times10^{-5}$ and
set $\rho=2\sigma$ \cite{mber2}. The step sizes for LMS adaptive
full-rank, SG adaptive MBER  full-rank and the other reduced-rank
techniques are $0.085$, $0.05$ and $0.035$, respectively. The
initial full-rank, reduced-rank  and  interpolation filters are
$[1,0,\ldots,0]^{T}$. The algorithms process $200$ symbols in TR and
$1000$ symbols in DD.

\begin{figure}[!htb]
\begin{center}
\def\epsfsize#1#2{1.0\columnwidth}
\hspace{-1em}\epsfbox{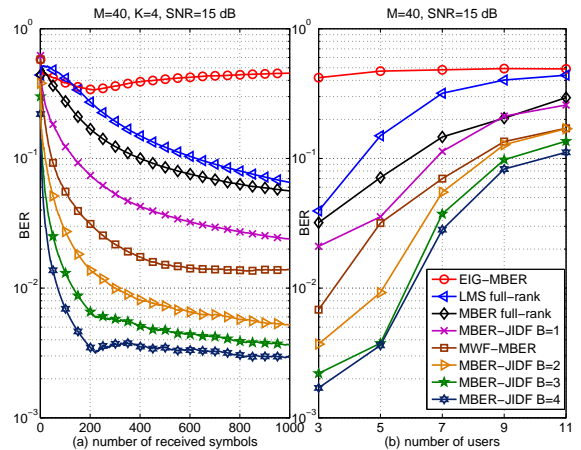} \vspace{-2em}\caption{BER
performance versus (a) number of received symbols, K$=4$, (b) number
of users. Parameters: $D=I=8$,  SNR$=15 $dB, and $N_{U}=2$.
}\label{fig:simulation1}
\end{center}
\end{figure}

Fig.\ref{fig:simulation1} (a) shows the BER performance of the
desired user versus the number of received symbols for the proposed
MBER-JIDF scheme and the conventional full rank and reduced-rank
algorithms. We set the rank $D=8$, $I=8$, $K=4$, $SNR=15$dB and
$f_{d}T=1\times 10^{-5}$. We can see that the proposed MBER-JIDF
reduced-rank algorithms converge much faster than the conventional
full rank and reduced-rank algorithms. Fig.\ref{fig:simulation1} (b)
illustrates the steady-state BER performance of the desired user
versus the number of users $K$. We can see that the best performance
is achieved by the proposed MBER-JIDF algorithms followed by the
MWF-MBER algorithm, the full-rank MBER algorithm, the full-rank LMS
algorithm and the eigen-decomposition-based algorithms. In
particular, the MBER-JIDF algorithm using $B=4$ can accommodate up
to four more users in comparison with the MWF-MBER algorithm
\cite{mber3}, at the BER level of $2\times 10^{-2}$.

\vspace{-1em}


%

%
\vspace{-2em}
\section{Conclusion}
\label{Section7:conclusion}

In this paper, we have proposed an adaptive reduced-rank linear
receive processing scheme and MBER algorithms for interference
suppression in large multiuser MIMO systems. For each branch, we
designed a group of jointly optimized interpolation and decimation
units, followed by linear receive filtering according to the
minimization of the BER cost function. The final output is switched
to the branch with the best performance based on the minimum error
probability. We have developed SG based algorithms for their
adaptive implementation. The results have shown that the proposed
scheme significantly outperforms existing algorithms and supports
systems with higher loads. Future work will consider non-linear
detectors, higher order modulation and other MIMO configurations.



\end{document}